\documentclass[amsmath,amssymbs,aps,twocolumn]{revtex4-2}

\usepackage{graphicx}
\usepackage{amsmath,amsfonts,amssymb}
\usepackage{epsfig}
\usepackage{wrapfig}
\usepackage{bbm}
\usepackage[usenames]{color}
\usepackage{array}
\usepackage{times}
\usepackage{float}
\usepackage{epstopdf}

\begin{document}

\preprint{APS/123-QED}

\title{Floquet Topological Dissipative Kerr Solitons and Incommensurate Frequency Combs}

\author{Seyed Danial Hashemi}

\author{Sunil Mittal}
\email[Email: ]{s.mittal@northeastern.edu}

\affiliation{Department of Electrical and Computer Engineering, Northeastern University, Boston, MA 02115, USA}
\affiliation{Institute for NanoSystems Innovation, Northeastern University, Boston, MA, 02115, USA}

\begin{abstract}
Generating coherent optical frequency combs in micro-ring resonators with Kerr nonlinearity has remarkably advanced the fundamental understanding and applications of temporal dissipative solitons. However, the spectrum of such soliton combs is restricted to the conventional definition of combs as phase-locked, equidistant lines in frequency. Here, we introduce a new class of floquet topological soliton combs that emerge in two-dimensional arrays of strongly coupled resonators engineered using floquet topology. Specifically, we demonstrate novel \textit{incommensurate} combs where the comb lines are not equidistant but remain phase-locked. These incommensurate combs are generated by self-organized, phase-locked floquet topological soliton molecules that circulate the edge of the array. We show that these floquet topological solitons are robust and they navigate around defects, allowing for agile tunability of the comb line spacing. Our results introduce a new paradigm in using floquet engineering to generate unconventional frequency combs beyond those achievable with single or weakly coupled resonators.
\end{abstract}

\maketitle

\noindent
\textbf{Introduction}\\
\noindent
The introduction of Kerr nonlinearity in resonators allows for the generation of dissipative Kerr temporal solitons that are pulses of light traveling in the resonator without any dispersion \cite{Cundiff2003, Firth1996, Kippenberg2011, Kippenberg2018, Gaeta2019}. This dispersion-free propagation is enabled by balancing the linear dispersion of the resonator against the dispersion induced by the Kerr nonlinearity, and the linear resonator losses against the nonlinear gain. In the frequency domain, such temporal solitons lead to the generation of coherent phase-locked optical frequency combs, called Kerr combs, that enable numerous applications, such as on-chip precision clocks, spectroscopy, on-chip microwave synthesis and processing, wavelength multiplexed optical transceivers, light detection and ranging (LiDARs), and so on \cite{Gaeta2019, Suh2016, Marin-Palomo2017, Spencer2018, Diddams2020, Riemensberger2020}. 

Recently, coupled-resonator systems have emerged as a paradigm to generate novel soliton states that are not accessible using single resonators \cite{Jang2018, Vasco2019, Helgason2021, Tikan2021, Boggio2022, Tusnin2023, Yuan2023}. For example, a photonic molecule, a system of two coupled resonators, has been used to explore the synchronization of solitons \cite{Jang2018}, to engineer dispersion \cite{Miller2015, Yuan2023}, and also to generate novel soliton states such as gear solitons \cite{Tikan2021}. Very recently, a two-dimensional (2D) array of coupled ring resonators has demonstrated the use of topological physics to generate a novel class of solitons, called the nested temporal solitons that are associated with nested optical frequency combs \cite{Mittal2021b, Flower2024}. Specifically, this 2D array creates a synthetic magnetic field for photons. This leads to the emergence of topological edge states that circulate the lattice boundary, forming a traveling-wave super-ring resonator. Pumping the edge states generates the nested comb - an oscillating set of edge state resonances, repeating every free-spectral range (FSR) of ring resonators. Within the 2D array, the nested comb is associated with the self-formation of nested temporal solitons - a burst of single-ring solitons circulating the edge of the lattice. 

Here, we introduce another novel class of topological frequency combs and temporal solitons, the floquet topological solitons that exist in strongly coupled 2D ring resonator arrays arranged as a square lattice. Because of strong coupling, the array exhibits the anomalous floquet phase where the topological edge states emerge even when all the bulk bands have a zero Chern number \cite{Afzal2018, Afzal2020, Liang2013, Pasek2014}. We show that pumping the edge states of this array generates a floquet comb that populates multiple edge bands interleaved by bulk bands. More importantly, this edge-bulk band interleaving leads to the formation of novel \textit{incommensurate combs} where all the comb lines, corresponding to oscillating edge states, are not equidistant, yet remarkably phase-locked. In the spatio-temporal domain, this phase-locked incommensurate comb is associated with the generation of novel super-soliton molecules that are phase-locked across multiple rings and circulate at the edge of the lattice. By tuning the system to another topological phase, the Chern insulator (CI) phase, where the bulk bands have non-zero Chern number, we demonstrate another novel floquet topological soliton state where there exists exactly one soliton in every alternating ring on the edge of the lattice, and all the individual solitons are phase-locked forming a large super-soliton molecule. In this state, the output comb spectrum populates only a single edge band and shows oscillation of only a single edge mode in each FSR. We show that these floquet soliton molecules circulating at the edge are robust and they route around defects in the lattice.  This routing increases the effective length of the lattice edge, and therefore, allows for post-fabrication agile tunability of the comb line spacing. These results pave the way for further investigations on the use of strongly coupled nonlinear resonator arrays, along with floquet and topological design principles \cite{Lu2014, Khanikaev2017, Ozawa2019, Smirnova2020, Price2022, Mehrabad2023, Rechtsman2013, Lindner2011}, to tailor dispersion, and thereby, achieve unconventional soliton states and optical frequency combs with engineered spectra.


\begin{figure*}
 \centering
 \includegraphics[width=0.98\textwidth]{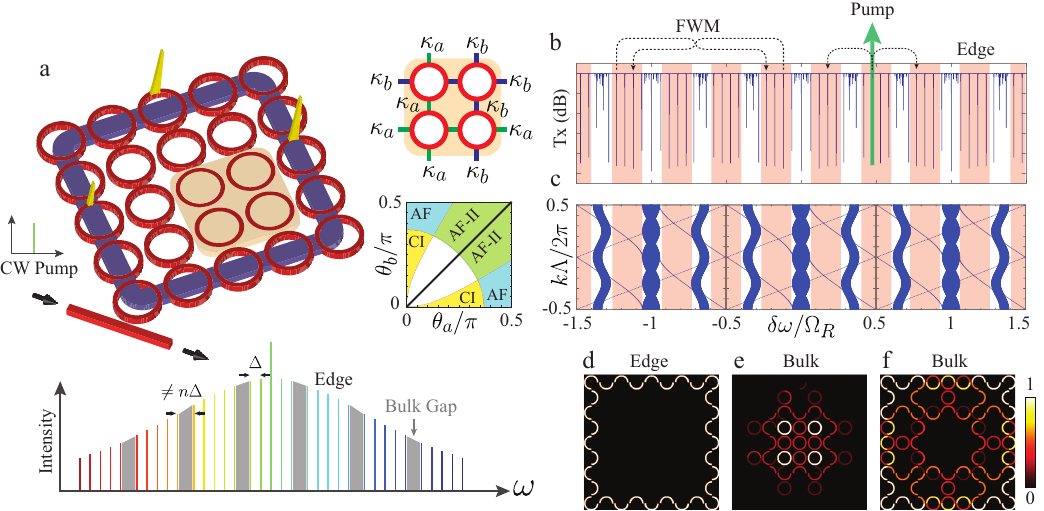}
 \caption{\textbf{a.} Schematic of the 2D ring resonator lattice that generates the floquet topological combs. The insets show a unit cell of the lattice (shaded), and the topological phase diagram for different coupling parameters $\kappa_{a}, ~ \kappa_{b}$.  The lattice is coupled to an input-output waveguide that facilitates the injection of a continuous-wave pump and out-coupling of the generated comb. The schematic also shows phase-locked soliton molecules in the lattice and the incommensurate comb spectrum. \textbf{b,c.} Transmission spectrum and band structures of the lattice for the anomalous floquet (AF) phase at $\theta_{A} = 0.45 \pi$, and $\theta_{B} = 0.05 \pi$. Edge bands are shaded red. The spectrum repeats every FSR of the ring resonators. The band structure is calculated for a semi-infinite lattice, with $k$ being the momentum and $\Lambda$ the periodicity of the lattice. \textbf{d-f} Intensity distribution in the lattice for the edge and the bulk states. The bulk state intensity distribution varies significantly with the pump frequency. 
}
 \label{fig:1}
\end{figure*}

\noindent
\textbf{Results}\\
\noindent
Our topological photonic system consists of a 2D square lattice of ring resonators (Fig.\ref{fig:1}a) \cite{Afzal2018, Afzal2020, Pasek2014}. The rings are evanescently coupled to their nearest neighbors. The coupling strengths between the resonators depend on their location in the lattice as shown in Fig.\ref{fig:1}a, and are parameterized as $\kappa_{a} = sin\left(\theta_{A}\right)$ and $\kappa_{b} = sin\left( \theta_{B}\right)$.  With this arrangement of coupling strengths, a unit cell of the lattice consists of four resonators. Accordingly, the lattice can host a maximum of four bands. However, the topology of these bands is dictated by the choice of coupling strengths $\kappa_{a}$ and $\kappa_{b}$.  In particular, the lattice exhibits non-trivial topology only when the coupling strength between the resonators is comparable to their FSR. In this regime of strong coupling, the system is referred to as a floquet system because it can not be described using an effective Hamiltonian and the single-mode approximation (close to a single FSR) that is typically used for describing weakly coupled ring resonator systems \cite{Hafezi2011, Hafezi2013, Leykam2018}. This resonator system can also be mapped to periodically modulated waveguide arrays implementing a periodically modulated Hamiltonian \cite{Afzal2018,  Maczewsky2017}. Furthermore, depending on the choice of coupling strengths, the lattice can host the anomalous floquet (AF) topological phase or the Chern insulator (CI) phase \cite{Afzal2018, Afzal2020}. 

\begin{figure*}
 \centering
 \includegraphics[width=0.98\textwidth]{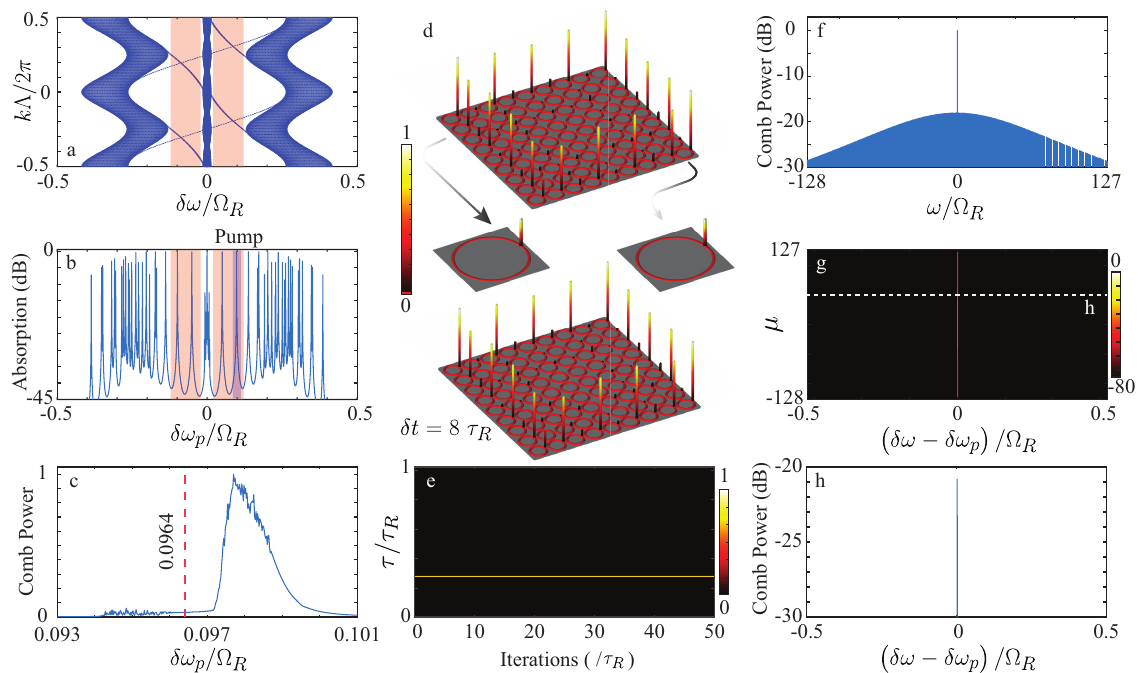}
 \caption {Generation of combs in the CI topological phase, with $\theta_{A} = 0.3\pi, ~ \theta_{B} = 0.01 \pi$. \textbf{a,b.} The band structure and the linear power absorption spectrum of the lattice show two edge bands and three bulk bands. \textbf{c.} Pump power in the input-output as a function of pump frequency. \textbf{d}. Intensity distributions in the lattice, at time instances separated by $\delta t = 8 \tau_{R}$, show the generation of a single soliton in each alternating ring at the edge of the lattice. This soliton molecule state appears to be stationary in the lattice. \textbf{e}. Comb output in the time domain shows pulses repeating every $\tau_{R}$. \textbf{f-i}. Comb output in the frequency domain shows oscillation of only a single edge mode (the pumped mode). }
 \label{fig:4}
\end{figure*}

In the AF phase, which exists, for example, at $\theta_{A} = 0.45 \pi, ~\theta_{B} = 0.05 \pi$, edge states appear in all band gaps, at normalized frequency detunings $\delta\omega = \left(\omega - \omega_{0}\right) \sim \pm 0.15 ~\Omega_{R}$ and at $\delta\omega \sim \pm 0.5 ~\Omega_{R}$ (Fig.\ref{fig:1}b, c). Here $\omega$ is the input light frequency, $\omega_{0}$ is the ring resonance frequency for a given longitudinal mode, and $\Omega_{R}$ is the free-spectral range of the rings (in angular frequency units). This topological phase is referred to as the anomalous phase because the edge states appear even though the Chern number for all the bands is zero \cite{Afzal2018, Pasek2014}. Such a topological phase has also been observed in floquet systems implemented using periodically modulated photonic waveguides \cite{Maczewsky2017}. In contrast, in the Chern insulator phase, which exists, for example, at $\theta_{A} = 0.3 \pi, ~\theta_{B} = 0.01 \pi$, the bulk bands have a non-zero Chern number. In this phase, the edge states appear in only two of the four bandgaps, at $\delta\omega \sim \pm 0.06 ~\Omega_{R}$ (Fig.\ref{fig:4}a,b). The array hosts another anomalous-floquet phase (the AF-II), for example, at $\theta_{A} = 0.45 \pi, ~\theta_{B} = 0.2 \pi$ where a small band gap supporting edge states appears at $\delta\omega \sim 0.5 ~\Omega_{R}$. We did not find any stable soliton solutions in this phase (see Supplementary Section IV). 

We emphasize that the band structure and the transmission/absorption spectrum repeat at every longitudinal mode resonance frequency of the ring resonators, spaced by the FSR $\Omega_{R}$ (Fig. \ref{fig:1}b,c). Furthermore, the topological phases, the widths of the bulk and the edge bands are dictated by the particular choice of the couplings $\kappa_{a}$ and $\kappa_{b}$. As we will show in the following, this property allows us to tune the generated comb spectra by tuning the coupling parameters. We also note that because of the strong coupling and the floquet nature of our system, the edge state resonances can appear at frequency detunings $\delta\omega = \left(\omega-\omega_{0}\right)$ that are comparable with the FSR $\Omega_{R}$. In this regime, the usual Lugiato-Lefever formalism that relies on the single-mode approximation for simulating the generation of Kerr combs is no longer valid \cite{Little1997, Chembo2013, Hansson2016}. 

\begin{figure*}
 \centering
 \includegraphics[width=0.98\textwidth]{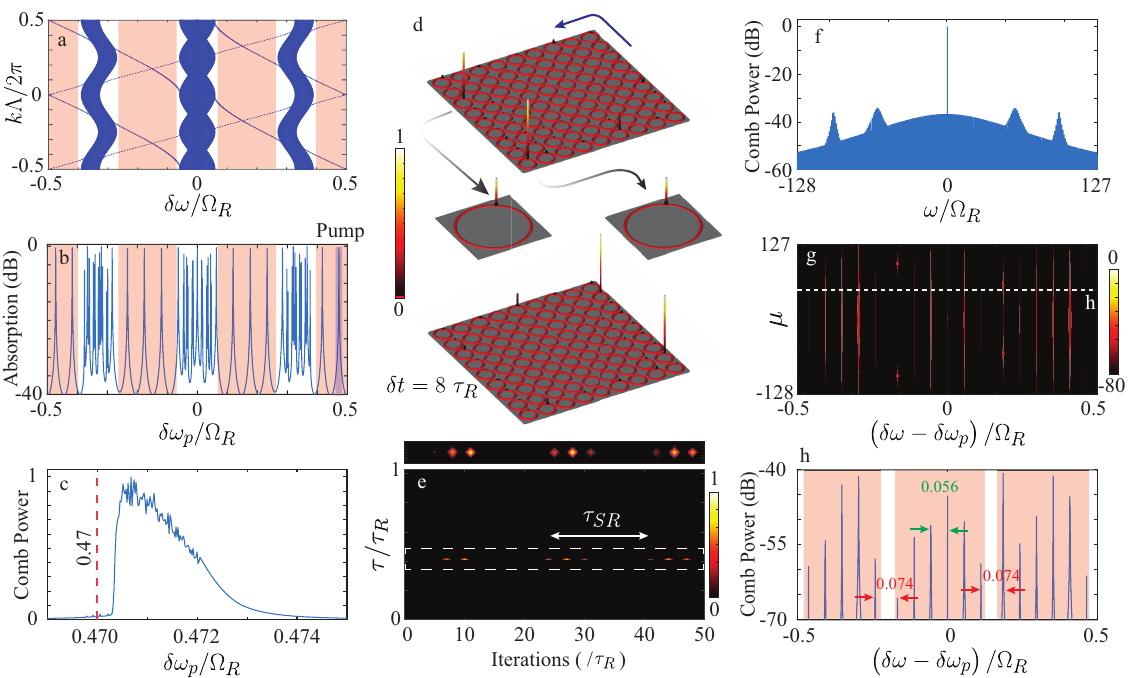}
 \caption {Generation of floquet solitons and incommensurate combs in the AF phase, for $\theta_{A} = 0.45 \pi, ~\theta_{B} = 0.05 \pi$. \textbf{a.} Band structure, \textbf{b.} absorption spectrum of the lattice, in the linear regime, showing edge and bulk bands. The band structure shown in \textbf{a.} is a zoom-in of Fig.\ref{fig:1}c within a single FSR. The absorption spectrum in \textbf{c.} is calculated as (1-Tx), with the transmission (Tx) shown in Fig.\ref{fig:1}b, and is zoomed-in within a single FSR. The pumped edge mode is highlighted in blue. \textbf{c.} Comb power (normalized) in the input-output ring as a function of pump frequency in the non-linear regime. The pump frequency where we observe coherent incommensurate combs soliton molecules is indicated. \textbf{d.} Intensity distribution in the lattice showing phase-locked super-soliton molecules. The super-solitons circulate the edge in the CCW direction as time evolves (here time interval $\delta t = 8 \tau_{R} $). Insets show the positions of solitons in the individual rings. To show the location of the rings, we have added a constant intensity (normalized) background of 0.01. \textbf{e.} Output of the floquet topological comb in the time domain, showing super-soliton pulses repeating after round trip time in the super-ring resonator. The inset shows a zoom-in of the temporal output where relative intensity variations in the output pulses are clearly visible. \textbf{f} Output comb spectrum. \textbf{g} Comb spectrum reorganized as a function of frequency detuning $\delta\omega/\Omega_{R}$ and FSR index $\mu$. \textbf{h} Comb spectrum showing oscillation of edge modes in a single FSR (here $\mu = 50$). The incommensurate nature of the comb is evident in \textbf{g, h} where the oscillating edge mode frequencies are not all equidistant.}
 \label{fig:2}
\end{figure*}

To generate optical frequency combs, the lattice is pumped using a continuous-wave laser through an input-output waveguide which is coupled to one of the lattice sites (Fig.\ref{fig:1}), with a coupling coefficient $\kappa_{\text{IO}}$. In the lattice, the pump laser generates the optical frequency comb via the four-wave mixing process, and the generated comb appears at the output port of the input-output waveguide. The nonlinear propagation of the fields inside the ring resonators and subsequent generation of combs is simulated using the Ikeda map approach detailed in the Supplementary Information (Section I). We use dimensionless parameters such that the total ring length $L_{R}$, the round-trip time $\tau_{R}$, and, consequently, the group velocity $v_{g}$ are all set to unity, and the FSR $\Omega_{R} = \frac{2\pi}{\tau_{R}} = 2\pi$. Similarly, the fields are normalized such that the strength of nonlinearity $\gamma$ is also set to unity. The resonator waveguides are assumed to have a loss coefficient $\alpha = 2\times 10^{-4}$. To generate bright solitons, we assume that the ring resonators have an anomalous dispersion, given by second-order dispersion parameter $D_{2}$ which we set to  $4\times10^{-6} ~ \Omega_{R}$. We emphasize that we do not assume any dispersion profile for the supermodes of the resonator array. The resonance frequencies of the supermodes, which include both the edge and the bulk states, emerge naturally from the simulation framework. Different soliton states are generated by an appropriate choice of the coupling parameters $\theta_{A}, ~\theta_{B}$, and by tuning the pump frequency and pump power.\\


\begin{figure*}
 \centering
 \includegraphics[width=0.98\textwidth]{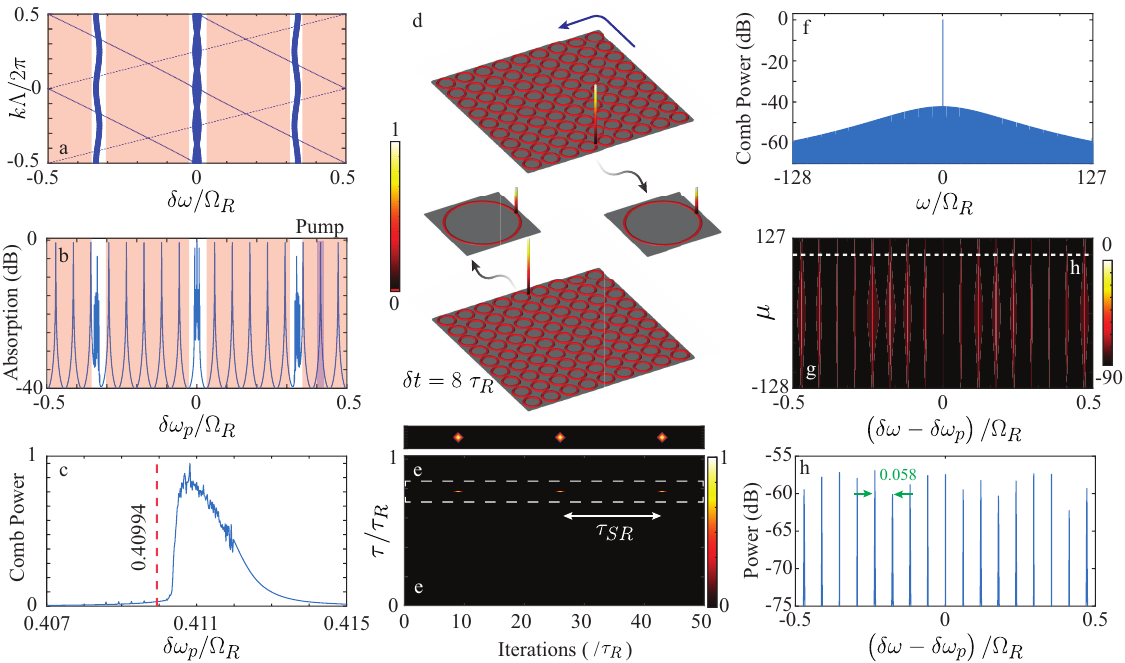}
 \caption {Generation of commensurate combs in the AF topological phase, with $\theta_{A} = 0.49\pi, ~ \theta_{B} = 0.01 \pi$. \textbf{a,b} The band structure and the linear power absorption spectrum of the lattice show reduced bulk band gaps. \textbf{c.} Comb power (normalized) in the input-output as a function of pump frequency. \textbf{d}. Intensity distributions in the lattice, at time instances separated by $\delta t = 8\tau_{R}$, show the generation of a single super-soliton in the lattice that circulates the edge in a counter-clockwise direction.    \textbf{e}. Comb output in the time domain shows the generation of a single pulse, repeating every $\tau_{SR}$. The inset shows a zoom-in of the temporal output. \textbf{f-g}. Comb output in the frequency domain shows a regular comb with all edge modes oscillating and no gaps.}
 \label{fig:3}
\end{figure*}

\noindent
\textbf{Incommensurate Combs  in the AF Phase}\\
\noindent
To demonstrate the generation of floquet topological solitons, we start with the anomalous floquet phase (AF), implemented in a $9 \times 9$ lattice of rings with $\theta_{A} = 0.45 \pi$, $\theta_{B} = 0.05 \pi$. The band structure for a semi-infinite lattice and the corresponding power absorption spectrum for the $9 \times 9$ lattice are shown in Fig.\ref{fig:2}a, b, respectively, as a function of the normalized pump frequency detuning $\delta\omega_{p} = \left(\omega_{p}-\omega_{0}\right)$, where $\omega_{p}$ is the input pump frequency. The edge states (highlighted in red) manifest as regularly spaced resonances (peaks) in the absorption spectrum. Because the edge states circulate around the complete boundary of the lattice, they constitute a super-ring resonator. Therefore, the edge state resonances represent the longitudinal modes of this super-ring resonator. In contrast to the edge states, the bulk mode resonances do not show any regular structure and their wavefunction occupies the bulk of the lattice (Fig.\ref{fig:1}e,f). 

We pump near one of the edge state resonances as shown in Fig.\ref{fig:2}b, with normalized pump field $E_{in} = 0.018$. The resulting nonlinear absorption as a function of the pump frequency is shown in Fig.\ref{fig:2}c. When the input pump frequency detuning $\delta\omega_{p} = 0.47 ~\Omega_{R}$, we observe the generation of a novel floquet super-soliton molecule at the edge of the lattice (Fig.\ref{fig:2}d). In this state, three different rings on the edge host a single soliton each. More importantly, we find that the three solitons are always phase-locked, that is, their relative positions in the rings are exactly the same. We refer to this state as a super-soliton molecule. As time evolves, this super-soliton molecule circulates around the edge in the counter-clockwise direction. A movie showing the evolution of this super-soliton molecule is available as Movie 1. Furthermore, because of the unique nature of the edge states, neither soliton completes a round-trip in the single ring. This is in contrast to the conventional single-ring solitons, where the soliton is defined as a non-dispersive pulse circulating in the ring resonator. We note that the relative intensities of the three solitons oscillate as a function of time, similar to that observed for breathing solitons in single-ring resonators \cite{Yu2017}. Nevertheless, we have confirmed that the total power in the super-soliton molecule stays constant. 

This generation of super-soliton molecules in the lattice manifests at the output port as a repeating set of three pulses (Fig.\ref{fig:2}e). The pulses appear at the same relative times $\tau$, indicating that they are phase-locked across all rings. Their repetition rate, equal to $\sim 17 \tau_{R}$, corresponds to the round-trip time $\tau_{SR}$ in the super-ring resonator formed by edge states. Similar to the intensity distributions within the lattice, the relative intensities of the three output pulses show oscillation from one period to the next. 

The generated frequency comb spectrum corresponding to this soliton molecule state is shown in Figs.\ref{fig:2}f-h. We observe that the intensity in the comb spectrum is smooth, indicating the existence of a coherent super-soliton state (Figs.\ref{fig:2}d,e). To better visualize the nature of oscillating modes in the comb, in Fig. \ref{fig:2}g, we plot the generated comb spectrum as a function of the longitudinal mode index $\mu$ and the normalized frequency detunings $\delta\omega - \delta\omega_{p} =  \left(\omega - \omega_{0,\mu}\right) - \left(\omega_{p} - \omega_{0,0} \right)$. Here $\omega$ is the frequency of the generated comb line, and $\omega_{0,\mu}$ is the resonance frequency for mode $\mu$, with $\mu = 0$ corresponding to the pumped FSR. Therefore, $\delta\omega - \delta\omega_{p} $ represents frequency detuning within one FSR.  

At each longitudinal mode index $\mu$, we observe the oscillation of multiple comb lines, which are indeed the edge state resonances (Fig.\ref{fig:2}g, h). More importantly, we find that the oscillating edge state resonances span the complete FSR, encompassing three edge bands (shaded red). Within each edge band, the comb lines are regularly spaced with a frequency spacing $\Omega_{SR} \simeq 0.056 \Omega_{R}$. However, the adjacent comb lines in different (neighboring) edge bands are spaced by $\simeq 0.074 \Omega_{R}$, which is not an integer multiple of $\Omega_{SR}$, the free-spectral range of the super-ring resonator. This clearly shows the generation of an \textit{incommensurate} floquet topological comb, which is phase-locked even when the constituent comb lines are not equidistant. We note that the periodicity of each oscillating edge mode (comb line) is exactly $\Omega_{R}$, the FSR of the single rings. Therefore, the observed incommensurate combs can be interpreted as sets of several (here 17) combs, each set corresponding to one oscillating edge mode and a line spacing $\Omega_{R}$. Different sets of combs have different frequency offsets (with in $\Omega_{R}$), and therefore, correspond to different carrier-envelope phase.  We do not observe any oscillation of the bulk modes in the spectrum, consistent with our observation of negligible intensity in the bulk of the lattice (Fig.\ref{fig:2}d).

In Fig.\ref{fig:2}f, we also observe the formation of satellite peaks in the incommensurate comb spectrum, similar to those observed in single-ring resonators in the presence of higher-order dispersion which leads to dispersive waves \cite{Kippenberg2018}. From Fig.\ref{fig:2}g, we find that these satellite peaks are associated with only two edge modes located at the boundary between the edge and the bulk bands. It is interesting to note that here, we assumed that the higher-order dispersion of the single-ring resonators is zero. Similarly, the edge states are generally assumed to have a linear dispersion in the edge band. However, the band structure in Fig.\ref{fig:2}a shows that the edge states do have a non-zero higher-order dispersion near the boundary between the edge and the bulk bands. This higher-order dispersion of edge states generates the satellite peaks. 

The anomalous floquet topological combs distinguish themselves from the nested topological combs observed in refs.\cite{Mittal2021b, Flower2024} where the oscillating edge state resonances populate a single edge band and are confined to a very close vicinity of the longitudinal mode resonance. Furthermore, unlike incommensurate combs, all the oscillating comb lines are regularly spaced in the nested topological combs. \\
\\
\noindent
\textbf{Commensurate Combs in the AF Phase}\\
\noindent
The incommensurate nature of the frequency combs observed here is because of the fact that the edge states participating in the comb formation are inter-spaced by bulk bands. This implies that by tuning the coupling parameters and, thereby, tuning the bulk bandwidth, it should be possible to generate regularly spaced commensurate combs. To show that this is indeed the case, in Fig.\ref{fig:3}, we show results for coupling parameters $\theta_{A} = 0.49 \pi$, and $\theta_{B} = 0.01 \pi$. This choice of coupling parameters still corresponds to the anomalous-floquet (AF) phase, but it significantly reduces the bulk bandwidths and, at the same time, increases the edge bandwidths. At a pump frequency detuning $\delta\omega_{p} = 0.40994 \Omega_{R}$, we now observe the formation of a single super-soliton in the lattice. In this state, only a single ring at the edge of the lattice hosts a single soliton that circulates the lattice in the counter-clockwise direction (Movie 2). Accordingly, the temporal output consists of a single pulse, repeating after the round-trip time $\tau_{SR} \sim 17 \tau_{R}$ of the super-ring resonator. The frequency comb spectrum, in this case, is regularly spaced and does not exhibit any incommensurate gaps. We also note the absence of satellite peaks in the comb spectrum, which is consistent with the observation of negligible higher-order dispersion of edge states in the band structure, as shown in Fig.\ref{fig:3}a. We note that other choices of the coupling parameters in the AF-phase could also yield similar commensurate combs as long as the bulk bandwidth is smaller than the frequency spacing between the edge state resonances.

\begin{figure*}
 \centering
 \includegraphics[width=0.98\textwidth]{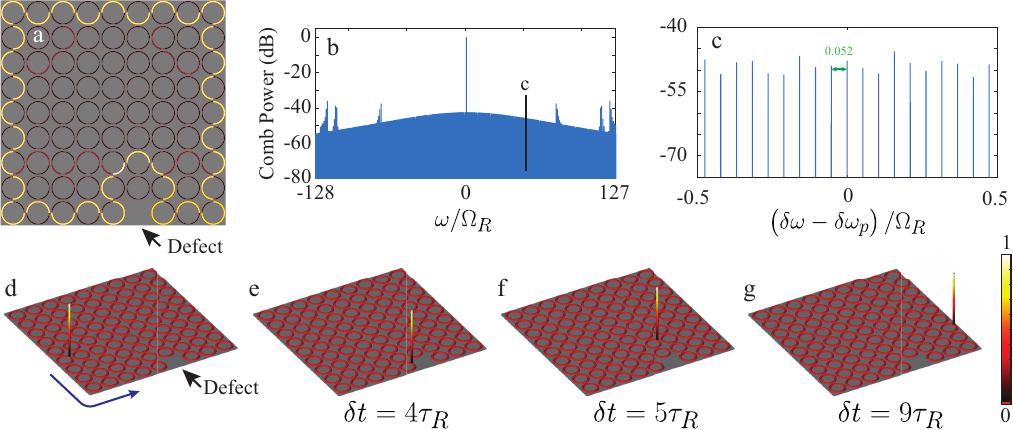}
 \caption {Robustness of edge states and tunability of comb line spacing. \textbf{a.} Routing of edge states around a defect in the lattice (in the linear regime). \textbf{b} Generated comb spectrum, and \textbf{c.} a zoom-in within a single FSR showing individual comb lines. \textbf{d-g} Routing of solitons around the defect without any scattering or loss of coherence.}
 \label{fig:5}
\end{figure*}

We emphasize that pumping the edge state resonances does not guarantee the formation of phase-locked solitons and coherent frequency combs. For example, pumping the edge states with a pump frequency of $\delta\omega_{p} = 0.4706 ~\Omega_{R}$ generates a comb operating in the modulation instability (chaotic) regime (see Supplementary Fig.S1). Even though this chaotic comb shows oscillation of multiple edge state resonances, the resonances are not phase-locked, and the resulting comb spectrum is not smooth. \\
\\
\noindent
\textbf{Soliton Molecules in the CI Phase}\\
\noindent
Next, we demonstrate the formation of another novel floquet soliton state in the Chern insulator phase (Fig.\ref{fig:4}). This phase exhibits three bulk bands that are topologically non-trivial, that is, they have a non-zero Chern number $\left(-1, 0, 1\right)$ \cite{Afzal2018}. The three bulk bands sandwich two edge bands (Fig.\ref{fig:4}a,b). When we tune the pump frequency near one of the edge state resonances, at $\delta\omega_{p} = 0.0964 ~\Omega_{R}$, and adjust the normalized input pump field $E_{in} = 0.025$, we observe the generation of a completely different soliton molecule state in the lattice: alternate rings on the edge host exactly one soliton, except for the rings on the corners that always host solitons (Fig. \ref{fig:4}d). Furthermore, the positions of the solitons in the rings are phase-locked. As time evolves, this intensity distribution appears to be stationary in the lattice (also see Movie 3).  Similar to the AF phase, the solitons in this phase do not complete a round trip in any single-ring resonator. Nevertheless, the solitons self-organize in alternate rings such that the path length difference between consecutive solitons is exactly equal to the circumference of a single ring resonator. 

The output temporal profile, in this case, consists of a series of pulses repeating at $\tau_{R}$, the round-trip time of a single ring resonator (Fig.\ref{fig:4}e). This behavior is consistent with the observation of soliton pulses in alternating rings such that their path-length difference equals a single resonator's circumference. We also note that this behavior is completely distinct from that observed in the AF phase, where a set of pulses repeats after the round-trip time $\tau_{SR}$ in the super-resonator (Fig. \ref{fig:3}e). In the frequency domain, we now observe the oscillation of only a single edge state resonance corresponding to the pumped mode (Fig.\ref{fig:4}f-h).  This observation is consistent with the observation of stationary intensity profile in the lattice (Fig.\ref{fig:4}d). 

The reduction in the number of oscillating edge modes is most likely due to the decrease in the topological edge bandwidth which decreases the number of edge state resonances and also increases the effective higher-order dispersion from the band edges. We also note the absence of any satellite peaks in the spectrum despite the edge states showing higher-order dispersion near the band edges (\ref{fig:4}a). This is because the oscillation of a single edge mode renders their dispersion irrelevant. Increasing the size of the lattice and thereby, increasing the number of edge state resonances in a given edge band could lead to the oscillation of multiple edge states resonances in the comb spectrum. 

It is instructive to compare the formation of this soliton molecule state in the CI phase to the formation of Turing rolls (see Supplementary Fig. S2). In the case of Turing rolls, we observe the formation of many pulses in each ring. In comparison, for the CI soliton molecules, there is only one pulse in each alternating ring. In both cases, only one edge mode oscillates in the comb spectrum. However, for Turing rolls, this oscillation is confined only to a few FSRs, whereas, for the CI soliton, a single edge mode oscillates in each FSR of the rings.\\
\\
\noindent
\textbf{Robustness and Tunability of Floquet Combs}\\
\noindent
The hallmark of topological edge states is their robustness against defects and disorders. To demonstrate that the floquet topological solitons and combs inherit this robustness even in the presence of such strong nonlinearity, in Fig.\ref{fig:5}, we show the formation of a single soliton in a defected 2D lattice where a ring resonator has been deliberately omitted. For this lattice we chose the AF phase with $\theta_{A} = 0.49$, and $\theta_{B} = 0.01$ to generate commensurate combs. We observe that the soliton follows a path that coincides with the edge states in the defected lattice. More importantly, we find that the soliton is robust, circulating around the defect without any back-scattering or loosing its coherence Fig.\ref{fig:5}(d-g) (also see Movie 4). 

Interestingly, robust routing around the defect increases the effective length of the edge states. This leads to a concomitant decrease in the comb line spacing (Fig.\ref{fig:5}(b,c). Specifically, without the defect, the comb line spacing for the commensurate comb was 0.058 FSR (Fig.\ref{fig:3}(h)). With the defect, the line spacing has reduced to 0.052 FSR, a change of $\sim 10\%$. Introducing additional defects on the edge would further decrease the comb line spacing. We note that such defects can be dynamically introduced in the lattice by, for example, integrating thermal heaters on the ring resonators that use thermo-optic effect to detune the ring resonance frequencies \cite{Mittal2016}. Therefore, the robustness of the edge states provides a convenient route for post-fabrication agile tuning of the comb line spacing. Such tuning is not accessible using single-resonator Kerr combs. \\


\noindent
\textbf{Discussion}\\
\noindent
To summarize, we have theoretically demonstrated the existence of a new class of floquet topological dissipative Kerr solitons and coherent combs in strongly coupled ring resonator arrays.  In particular, the incommensurate combs observed here go far beyond the conventional definition of frequency combs as a set of equidistant frequencies. Our demonstration builds on the now mature, coupled ring resonator platform that has enabled, for example, the observation of topological edge states of light \cite{Hafezi2013, Afzal2020}, their robustness against fabrication imperfections\cite{Mittal2014}, observation of higher-order corner states \cite{Mittal2019b}, implementation of topological lasers \cite{Bandres2018}, quantum light sources \cite{Mittal2018, Mittal2021, Dai2022, Afzal2023}, and non-Hermitian light steering \cite{Zhao2019}. The range of parameters, such as the loss, dispersion, coupling ratios, etc., used for our simulations is consistent with that used typically for single-resonator Kerr combs and can be achieved using the commercial low-loss silicon-nitride platform (see SI Section I). These floquet topological soliton combs augment the recently observed topological nested combs that operate in the weak coupling regime \cite{Mittal2021b, Flower2024}.

On a fundamental level, the floquet topological solitons observed here are temporal counterparts of floquet topological spatial solitons that have been recently observed in coupled waveguide arrays \cite{Mukherjee2020, Mukherjee2021, Maczewsky2020}. Therefore, our demonstration could pave the way for further exploring intriguing physics at the intersection of optical nonlinearity, topology, and synthetic dimensions and realizing, for example, a temporal analog of the quantized topological pumping of spatial solitons \cite{Jurgensen2021, Mostaan2022, Jurgensen2023, Smirnova2020}.

\vspace{12pt}
\noindent
\textbf{Acknowledgements}
This research was supported by startup and TIER 1 grants from Northeastern University. \\

\noindent
\textbf{Author Contributions:} S.M. conceived the idea and developed the simulation framework. S.D.H. performed the numerical simulations. Both authors contributed to analyzing the data and writing the manuscript. S.M. supervised the project.\\

\noindent
\textbf{Competing interests:} S.M. and S.D.H. have filed an invention disclosure based on the results reported in this manuscript.\\

\bibliographystyle{NatureMag}
\bibliography{FC_Biblio, Topo_Biblio}


\end{document}


\preprint{APS/123-QED}

\title{Supplementary Information: Floquet Topological Dissipative Kerr Solitons and Incommensurate Frequency Combs}

\author{Seyed Danial Hashemi}

\author{Sunil Mittal}
\email[Email: ]{s.mittal@northeastern.edu}

\affiliation{Department of Electrical and Computer Engineering, Northeastern University, Boston, MA 02115, USA}
\affiliation{Institute for NanoSystems Innovation, Northeastern University, Boston, MA, 02115, USA}

\maketitle

\section{Simulation Framework} 

We use the Ikeda map approach to simulate the nonlinear propagation of optical fields, and subsequent generation of optical frequency combs in the ring resonator array. The Ikeda map equations describing the propagation of the field in resonators, away from coupling regions, is written as \cite{Ikeda1979, Hansson2015, Hansson2016}
\begin{equation}
\frac{dE_{x,y}^{m}\left(z,\mu\right)}{dz} =  i\frac{2\pi}{L_{R}\Omega_{R}}\left(\omega_{p} + \Omega_{R} \mu - \omega_{0\mu} \right) E_{x,y}^{m}\left(z,\mu\right) - \frac{\alpha}{2}  E_{x,y}^{m}\left(z,\mu\right) + i \gamma \frac{1}{\tau_{R}} \int_{0}^{\tau_{R}} d\tau \left|E_{x,y}^{m}\left(z,\tau\right)\right|^{2} E_{x,y}^{m}\left(z,\tau\right) e^{i\omega_{\mu} \tau}.
\end{equation}
Here $E_{x,y}^{m}\left(z,\tau\right)$ is the field in a resonator at location $\left(x,y \right)$, at position $z$ and time $\tau = \left\{0,\tau_{R}\right\}$, with $\tau_{R}$ being the round-trip time in the ring. Its Fourier transform $E_{x,y}^{m}\left(z,\mu\right)$ is the field in the frequency domain with $\mu$ indicating the corresponding longitudinal mode index. The subscript $m$ labels the round-trip number. $\alpha$ is the propagation loss. The dispersion of the ring resonators is included in their resonance frequencies for a given FSR, indexed by $\mu$, as
\begin{equation}
 \omega_{0,\mu} = \omega_{0} + \Omega_{R} ~\mu + \frac{D_{2}}{2} \mu^{2},
\end{equation}
where $\Omega_{R} = 2\pi \frac{v_{g}}{L_{R}}$ is the free spectral range (in angular frequency units), $L_{R}$ is the length of the ring resonator, and  $D_{2}$ is the second-order dispersion which we assume to be anomalous. $\gamma = 2 \epsilon_{0} n_{0} n_{2} \omega_{0} $ is the strength of Kerr nonlinearity \cite{NLOptics_Boyd}. In this form of the equation, linear evolution is described in the frequency domain, and nonlinear evolution is described in the time domain, which permits using a split-step process for its numerical solution. 

To further simplify the above equation, we use dimensionless parameters and normalized coordinates such that $z ~\rightarrow ~\frac{z}{L_{R}} = \left\{ 0,1 \right\}$. Similarly, $\tau ~\rightarrow ~\frac{\tau}{\tau_{R}} = \left\{0,1 \right\}$. This also implies that the dimensionless FSR  $\Omega_{R} = 2\pi$, and the group velocity $v_{g} = 1$. The dimensionless fields are then scaled as $E_{x,y}^{m}\left(z,\tau\right) ~\rightarrow ~\sqrt{\gamma ~L} E_{x,y}^{m}\left(z,\tau\right)$ which effectively sets the strength of Kerr nonlinearity $\gamma = 1$. Using these dimensionless parameters, the simplified propagation equation is written as
\begin{equation}
    \frac{dE_{x,y}^{m}\left(z,\mu\right)}{dz} =  i\left(\omega_{p} - \omega_{0} - \frac{D_{2}}{2} \mu^{2} \right) E_{x,y}^{m}\left(z,\mu\right) - \frac{\alpha}{2}  E_{x,y}^{m}\left(z,\mu\right) + i \int_{0}^{1} d\tau \left|E_{x,y}^{m}\left(z,\tau\right)\right|^{2} E_{x,y}^{m}\left(z,\tau\right) e^{i\omega_{\mu} \tau}.
\end{equation}

Note that this equation is written in a frame moving at velocity $v_{g}$ in the ring resonators and features no dependence on $\Omega_{R}$. Therefore, a soliton pulse, moving at velocity $v_{g}$, would appear stationary.

The coupling between the rings is described using $2\times 2$ beam-splitter matrices of the form
\begin{equation}
\begin{pmatrix} E_{x,y}^{m}\left(z_{c}, \tau \right)\\ E_{x',y'}^{m}\left(z_{c},\tau \right) \end{pmatrix} = \begin{pmatrix} &t &i\kappa \\ &i\kappa &t\end{pmatrix} \begin{pmatrix} E_{x,y}^{m}\left(z_{c},\tau \right)\\ E_{x',y'}^{m}\left(z_{c},\tau \right) \end{pmatrix}.
\end{equation}
Here $t = \sqrt{1-\kappa^2}$,  $\kappa = \left(\kappa{a},\kappa_{b} \right)$ is the coupling coefficient between rings indexed by $\left(x,y \right)$ and its nearest neighbor $\left(x',y' \right)$, at coupling location indicated by $z_{c}$. The coupling of the input-output waveguide to the lattice at location $\left(x,y\right) = \left(1,1\right)$ is similarly described, using coupling coefficient $\kappa_{IO}$, such that 
\begin{equation}
 E_{out}^{m}\left(\tau\right) = i\kappa_{IO} ~E_{1,1}^{m}\left(z=1, \tau \right) + t_{IO} E_{in} \left(\tau \right).
\end{equation}
Here, $E_{in} \left(\tau \right)$ is the input pump field, and $E_{out}^{m}\left(\tau\right)$ is the output field, at the $m^{\text{th}}$ iteration and at time $\tau$. This field yields the temporal output of the generated frequency comb at any time $t = \left(m-1\right) \tau_{R} + \tau$, and its Fourier transform yields the frequency spectrum $E_{out}\left(\omega\right)$. 

For our simulations, we chose $\alpha = 2\times 10^{-4}$, $D_{2} / \left(2\pi \right) = 4 \times 10^{-6}$, $t_{IO} = 0.985$ for combs in the anomalous floquet phase, $t_{IO} = 0.995$ for solitons in the CI phase, and $t_{IO} = 0.975$ for Turing rolls in the CI phase . The couplings $\kappa$ are site-dependent and are chosen to implement a given topological phase. For a ring resonator, say with an FSR $\Omega_{R}/\left(2\pi\right) = 250 ~\text{GHz}$, the dispersion $D_{2}/\left(2\pi\right) = 1 ~\text{MHz}$ and the intrinsic loss rate $\kappa_{in}/\left(2\pi\right) = 25 ~\text{MHz}$ which corresponds to an intrinsic quality factor $Q_{i} \sim 8 \times 10^{6}$. This range of parameters is accessible experimentally using low-loss photonic platforms such as silicon-nitride \cite{Tikan2021}. For this choice of parameters and the normalized input pump field $E_{in} = 0.018$ at which we observe soliton molecules in the AF phase, the pump power required to generate incommensurate soliton combs $P = \frac{c}{n_{2} \omega_{0} L_{R}} A_{eff} \left|E_{in} \right|^{2} \sim 0.6 ~W$, which is also typical for generating Kerr combs.

\section{Chaotic (Modulation Instability) Combs in Anomalous Floquet Insulator Phase}

\begin{figure*}[!ht]
 \centering
 \includegraphics[width=0.98\textwidth]{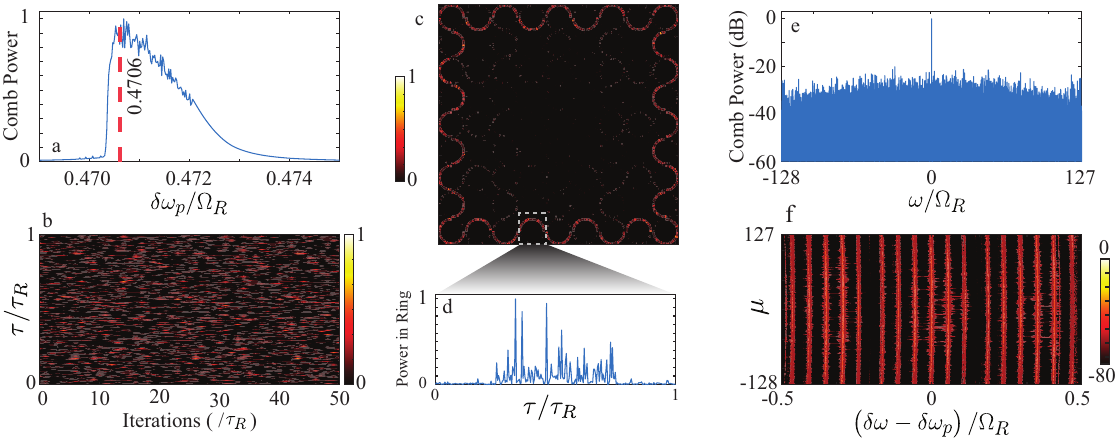}
 \caption{Chaotic Combs in the AF phase. \textbf{a.} Pump power in the super-ring. \textbf{b.} Temporal output shows no periodic patterns.  \textbf{c.} Intensity distribution in the lattice shows that the generated comb is still confined predominantly to the edge of the lattice. Intensities within a ring on the edge is shown in \textbf{d}. \textbf{e,f.} Generated comb spectrum shows oscillation of all the edge modes.}
 \label{fig:S1}
\end{figure*}

As we mentioned in the main text, pumping an edge state resonance does not guarantee the formation of soliton molecules and coherent optical frequency combs. A coherent comb is achieved by tuning the pump power and frequency. At most other pump powers and frequencies, we observe the formation of chaotic (modulation-instability) combs. The generated comb spectra and the lattice intensity profile for one such chaotic comb, in the AF phase, is shown in Fig.\ref{fig:S1}. This comb is generated at a different pump frequency detuning $\left(\omega_{p} =  0.4706 \right)$, but all the other parameters are same as in Fig. 2 of the main text. 

For the chaotic comb, as expected, the temporal output is chaotic with no periodic patterns. This is consistent with the observed intensity distribution within the lattice, which shows random variations within a ring and across different rings. Interestingly, the comb spectrum shows oscillation of all edge modes. Nevertheless, the edge modes are not phase-locked, and the underlying quadratic dispersion of the ring resonators can be observed in the curvature of the comb lines as a function of longitudinal mode index $\mu$. For the coherent soliton state (Fig. 2 of the main text), this dispersion is exactly balanced by the Kerr nonlinearity, and the oscillating comb lines show no curvature. 

\section{Turing Rolls in Chern Insulator Phase}

In addition to solitons, Kerr combs feature another coherent state called Turing rolls, where the spatio-temporal intensity profile in the ring exhibits many regularly spaced pulses, and the comb spectrum shows only a few oscillating longitudinal modes. The number of pulses in the ring matches the number of FSRs between oscillating modes.  

\begin{figure*} [!ht]
 \centering
 \includegraphics[width=0.98\textwidth]{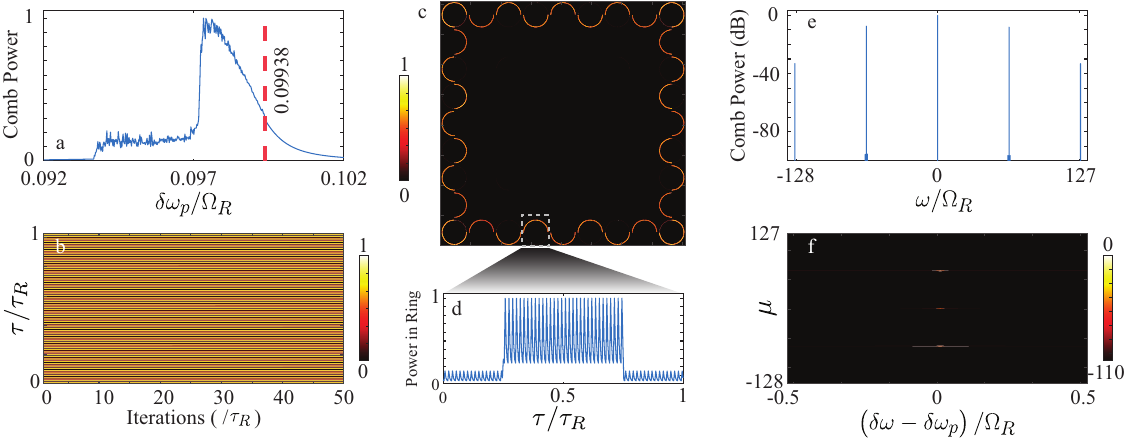}
 \caption{Turing rolls in the CI phase. \textbf{a.} Comb power in the IO ring as a function of pump frequency. The dashed line indicates the pump frequency where we observe Turing rolls. \textbf{b.} Output in the time domain shows pulses, repeating at a rate much faster than $\tau_{R}$. \textbf{c.} The intensity distribution in the lattice shows phase-locked Turing rolls in all the rings on the edge of the lattice. \textbf{d.} Intensity profile in a single ring resonator showing Turing rolls. \textbf{e,f} The comb spectrum shows oscillation of only a few longitudinal modes, with only one edge mode oscillating in each FSR.}
 \label{fig:S2}
\end{figure*}

We have also observed Turing rolls in the Chern insulator phase. The intensity distribution and the generated comb spectra are shown in Fig.\ref{fig:S3}. As before, we pump one of the edge states (shown in Fig. 4\textbf{b}). When $\delta\omega_{p} = 0.09938 \Omega_{R}$, we observe the formation of phase-locked Turing rolls in each ring on the edge of the lattice (Fig.\ref{fig:S2}\textbf{c}), that is, within each ring, we observe the formation of multiple regularly spaced pulses. Similar to the intensity distribution of the edge states in the linear case (Fig.1a), the intensity of Turing rolls is appreciable only in one-half of the ring resonators on the edge and three-fourths of the rings on the corner. The generated comb spectrum, shown in Fig.\ref{fig:S3}\textbf{e,f}, shows oscillation of only 4 longitudinal modes (in addition to the pumped FSR), separated by about 64 FSRs. This spacing matches the number of Turing rolls in each ring. Furthermore, within each FSR, only a single edge mode, which corresponds to the pumped mode, is populated. The intensity distribution (Fig.\ref{fig:S2}\textbf{b}) in the rings appears stationary because the simulations are carried out in a frame moving at velocity $v_{g}$.

The generation of Turing rolls in the CI phase can be compared to that of soliton molecules, where there is only a single soliton pulse in each alternating ring on the edge, and the comb spectrum shows oscillation of all longitudinal modes. 

\section{Frequency Combs in Anomalous Floquet - II Phase}

\begin{figure*} [!ht]
 \centering
 \includegraphics[width=0.9\textwidth]{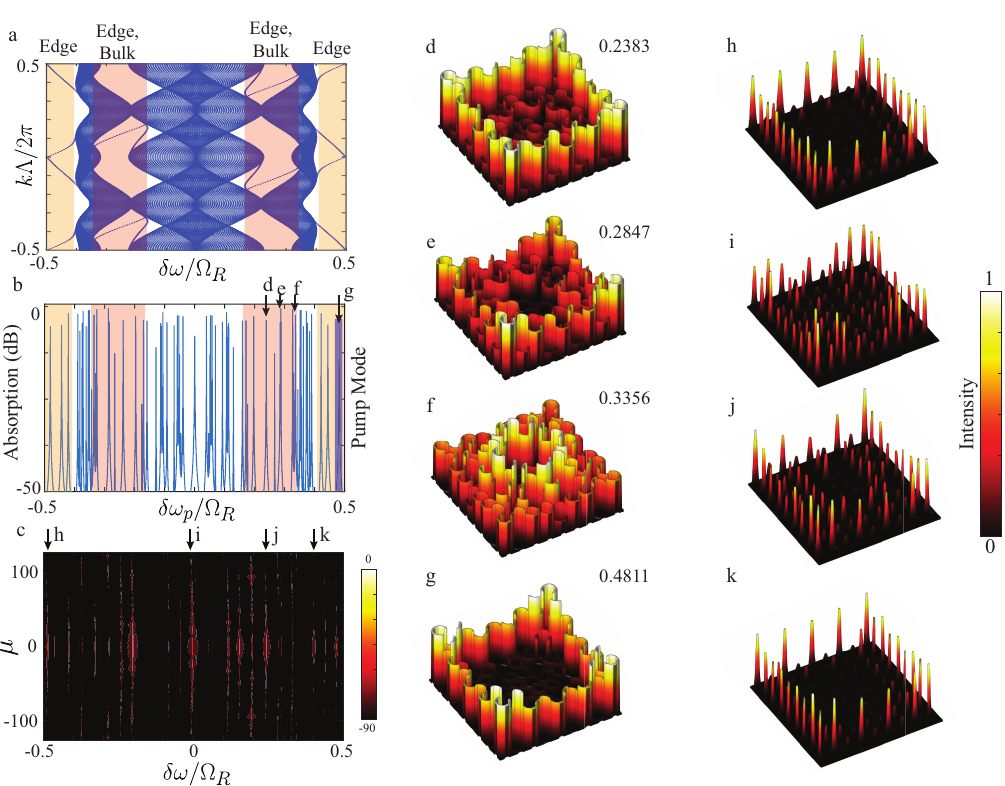}
 \caption{\textbf{a.} Band structure, and \textbf{b.} linear absorption spectrum for the AF-II phase, at $\theta_{A} = 0.45\pi, ~\theta_{B} = 0.2\pi$. \textbf{c.} Generated comb spectrum. \textbf{d-g} Lattice intensity profile, in the linear domain, for the resonances indicated in \textbf{b}. The lattice hosts edge and bulk states in the same band. \textbf{h-k} Lattice intensity distributions corresponding to the generated comb lines indicated in \textbf{c}. Pumping the edge state shows coupling to the bulk states.}
 \label{fig:S3}
\end{figure*}

In addtition to the anomalous flouqet (AF) and Chern insulator (CI) phases, our system exhibits another topological phase called the anomalous-floquet-II (AF-II) phase (Fig. 1 of the main text).  The band struture of the AF-II phase, for example, at $\theta_{A} = 0.45\pi, ~\theta_{B} = 0.2\pi$, exhibits edge states in two complete bandgaps (near $\delta\omega \sim 0.5 \Omega$) and two incomplete bandgaps (near $\delta\omega \sim 0.25 \Omega$) where the edge states live closer to the the bulk states (Fig.\ref{fig:S3}a). This arrangement is also evident in the transmission spectrum (Fig.\ref{fig:S3}b) and the corresponding intensity distributions in the finite lattice (Fig.\ref{fig:S3}d-g). For the edge state resonances in the complete band gaps ($\delta\omega \sim 0.48 ~\Omega_{R}$), the intensity distribution is mostly confined to the edge, with negligible intensity in the bulk (Fig.\ref{fig:S3}g). However, for resonances in the in-complete bandgap ( ($\delta\omega = 0.2383, 0.2847, 0.3356  ~\Omega_{R}$), the intensity distribution occupies both the edge and the bulk of the lattice

When pumping the edge states, we did not observe any stable coherent soliton solutions. We believe this is because of the presence of incomplete bandgap that simultaneously supports both edge and bulk states. Specifically, we observed that pumping the edge states in the complete bandgap region ($\delta\omega \sim 0.48 ~\Omega_{R}$) generates comb lines at bulk state resonances as can be confirmed by the lattice intensity distributions corresponding to generated comb lines (Fig.\ref{fig:S3}h-k). The observed edge-bulk coupling is mediated by the edge states of the incomplete band gap region that have substantial spatial overlap with the pumped edge states, and it prevents the formation of stable coherent solitons. 

\section{Alignment between Band Structures and Comb Spectra}

\begin{figure*} [!ht]
 \centering
 \includegraphics[width=0.88\textwidth]{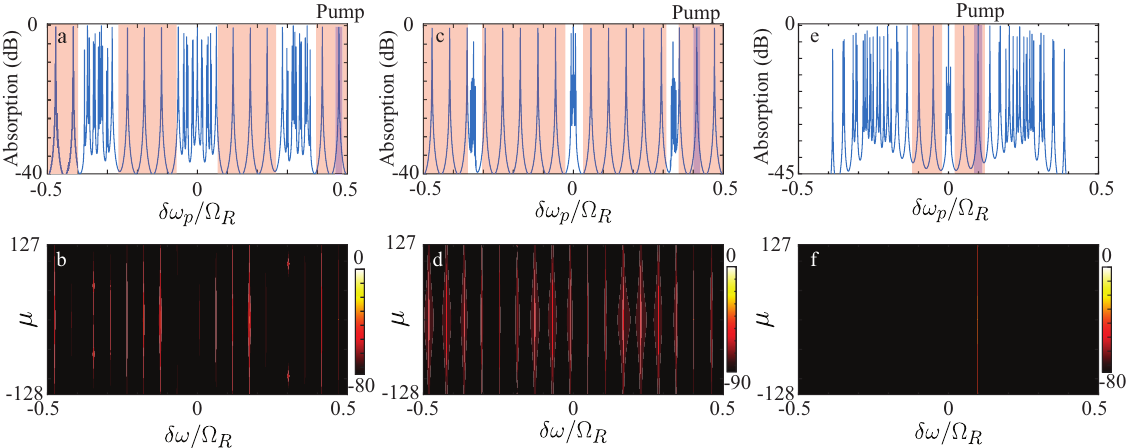}
 \caption{\textbf{a,b.} Linear absorption spectrum and observed comb spectrum for the AF-phase, in the incommensurate comb regime. \textbf{c,d}, and \textbf{e,f}, same for the AF-phase with commensurate spectrum, and the CI-phase. The frequency axis ($x-$axis) of these comb spectra is not normalized for the pump frequency.}
 \label{fig:S4}
\end{figure*}

In the main text, to show the symmetry in the comb generation, we presented comb spectra (Fig.2g, 3g, and 4g) that were normalized to the pump frequency. In Fig.\ref{fig:S4}, we show these comb spectra without the pump frequency normalization. As expected, we observe correspondence between the oscillating edge modes of the comb spectra and those of the linear transmission spectra. The bulk bands of the linear spectra manifest as gaps in the comb spectra.
  
\bibliographystyle{NatureMag}
\bibliography{FC_Biblio, Topo_Biblio}